# Adaptive link dynamics drive online hate networks and their mainstream influence

Minzhang Zheng[1,*], Richard Sear[1,*], Lucia Illari[1,*], Nicholas J. Restrepo[2], Neil F. Johnson[1,*]

[1]Dynamic Online Networks Laboratory, George Washington University, Washington, D.C., 20052, U.S.A.
[2]ClustrX LLC, Washington, D.C., U.S.A.
* These authors contributed equally to the paper

**Online hate is dynamic, adaptive**[1,2,3,4,5] **and is now surging**[6,7,8] **armed with AI/GPT tools**[9,10]**. Its consequences include personal traumas**[11,12,13]**, child sex abuse**[14,15] **and violent mass attacks**[16]**. Overcoming it will require knowing how it operates at scale**[17]**. Here we present this missing science and show that it contradicts current thinking. Waves of adaptive links connect the hate user base over time across a sea of smaller platforms, allowing hate networks to steadily strengthen, bypass mitigations, and increase their direct influence on the massive neighboring mainstream. The data suggests 1 in 10 of the global population have recently been exposed, including children. We provide governing dynamical equations derived from first principles. A tipping-point condition predicts more frequent future surges in content transmission. Using the U.S. Capitol attack and a 2023 mass shooting as illustrations, we show our findings provide abiding insights and quantitative predictions down to the hourly scale. The expected impacts of proposed mitigations can now be reliably predicted for the first time.**

Nearly 50% of all Americans now compromise aspects of their and their childrens' daily lives in order to lower the risk of experiencing some hate-driven mass shooting, e.g. 6 May 2023 Allen, Texas shooting which appears to be one of an increasing number inspired by social media hate content[18,19]. Separately, 2024 will see more than 60 elections across 54 countries including the U.S. and India, where the scope for online hate to cause voter intimidation is huge[20,21]. Such mass-scale threats, now supercharged by AI/GPT weaponry, are accelerating efforts to win the war against online hate and other harms[22,23,24,25,26,27,28,29,30,31,32,33,34,35,36,37]. The reviews in Refs. [38,39,40] offer unifying perspectives on this huge and still growing body of research, while Ref. [41] provides daily updates on new studies.

This war against online hate of all forms, is being led on the regulatory side by the EU's "Digital Services Act" (DSA) and "A.I. Act" [42,43]. Platforms on the list of "Very Large Online Platforms" such as Facebook and Twitter, must carry out a risk assessment which includes an analysis of how harmful content might be disseminated through their service[44]. At face value, this appears to make perfect sense since the largest platforms (e.g. Facebook, Twitter) have the largest share of users. Hate etc. are thought to occupy the fringes of the Internet[45,46,47,48,49,50]. However, winning any war requires an accurate picture of the battlefield. Here we build the most accurate picture to date, and we show how its dynamical features contradict such current thinking.

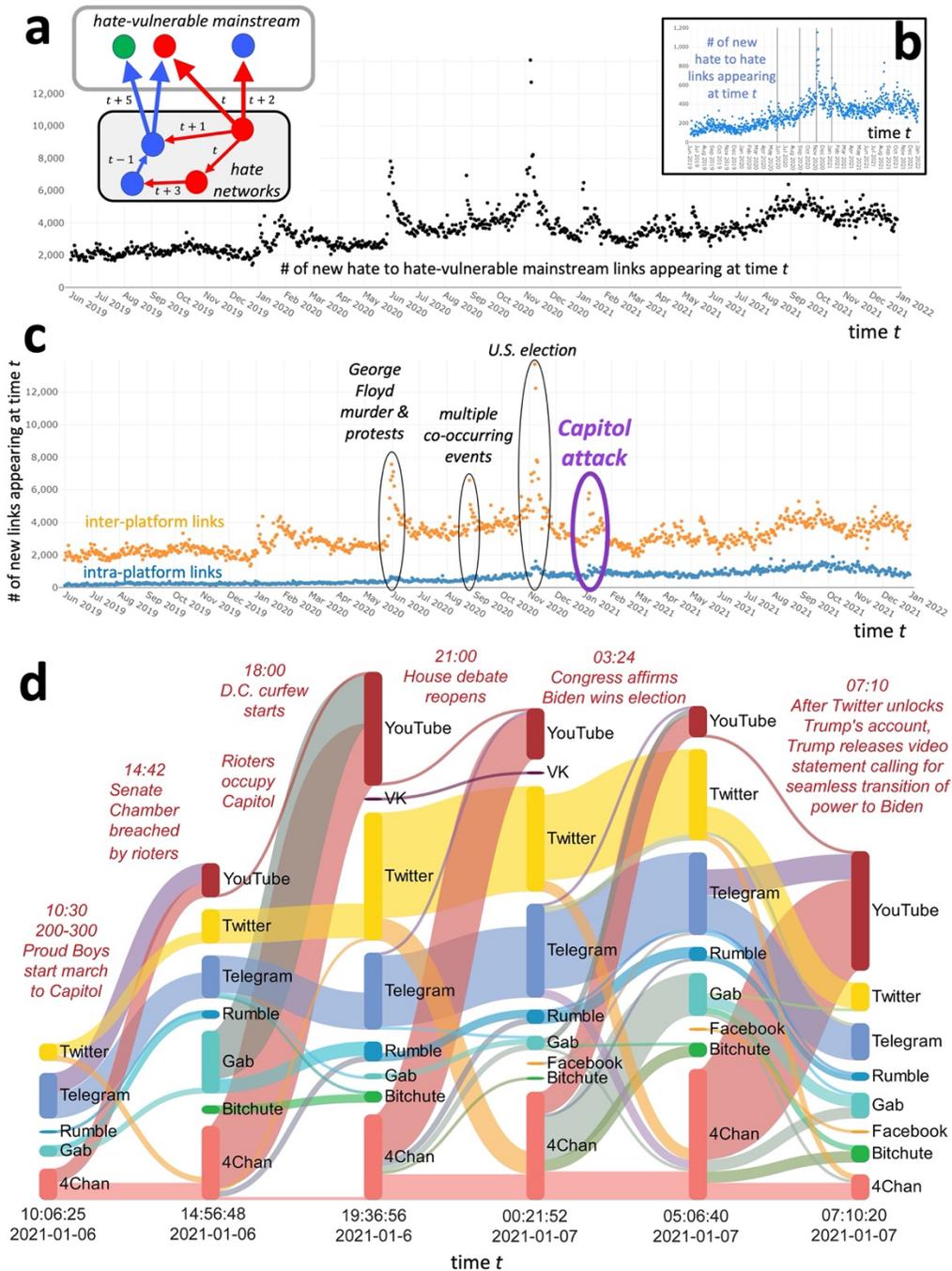

**Figure 1. Hate's highly adaptive link dynamics. a-c:** Number of new links created on each day *t* from hate communities (nodes). Time-series show different aggregations across the 4 link types (schematic in panel a left inset): hate to hate inter-platform; hate to hate intra-platform; hate to hate-vulnerable mainstream inter-platform; hate to hate-vulnerable mainstream intra-platform. SI Sec. 2 gives explicit examples of these links. **d:** Sankey diagram shows intra-day flows of new links from hate nodes on a given platform (source) into hate nodes on a target platform. SI Sec. 4 explains Sankey diagram construction.

Our methodology follows Refs. [51,52] (SI Sec. 1) but goes beyond prior studies by (i) including the hate-vulnerable mainstream communities that hate communities link to over time, (ii) tracking this data down to second-scale resolution across 13 platforms, (iii) including new decentralized[53] and blockchain platforms (e.g. Minds, Steemit) for which blame cannot be pinned on single servers and cryptocurrency can incentivize users, and (iv) including gaming-related platforms[54] such as Discord which played a key role in recent security leaks. Our focus is on in-built communities because people join these to develop their shared interests[55,56,57,58], including hate. Examples are a VKontakte Club (VKontakte is a social media platform controled by Russian state-owned bank Gazprombank and insurance company Sogaz); a Facebook Page; a Telegram Channel; a Gab Group. Each contains anywhere from a few to a few million users and is unrelated to network community detection. A "hate" community is one in which 2 or more of its 20 most recent posts include U.S. Department of Justice-defined hate speech. A "hate-vulnerable" community is one that is outside this hate community core, but was linked to directly by a hate community (Fig. 1a inset). Hate-vulnerable communities' views can vary significantly, but they mostly represent a benign mainstream that have become targets of the hate core (SI Sec. 1.2). A link to community B can appear in community A at some time $t$ if B's content is of interest to A's members (SI Figs. S1, S5 show examples). The link directs attention of A's members to B, which may be on a different platform and another language. A's members can then add comments and content on B without B's members knowing this incoming link exists. Hence, B's members can unwittingly experience direct exposure to, and influence from, A's hateful narratives. Since our focus is on hate networks, we do not include links originating in hate-vulnerable nodes. No individual information is required. Only public communities are accessed, but the ecosystem of open communities provides a skeleton on which private communities sit (Fig. S4).

The resulting battlefield (schematic Fig. 1a left inset; results Figs. 1-3; data in SI) is a directed dynamical network with link wiring that can change quickly over time within and across platforms, and strong direct linkage from the hate networks to the massive hate-vulnerable mainstream. It contains 1,848 hate communities (nodes) totaling roughly 25M individuals, that in 2.5 years have created 340,246 new links to each other and 2,899,115 links into 404,416 hate-vulnerable mainstream communities. While the nodes (communities) are fairly constant over time, the link number increases massively every day and hence steadily strengthens the hate networks and their potential mainstream influence. Each new link (e.g. Figs. S1,S5) means members of the source hate community (node) can immediately engage with the target community (node), pass hate content to it, and influence it.

Three key features emerge (illustrated in Figs. 1-3) that hate mitigations and legislation must account for in order to be effective:

(1) They must outpace the link creation dynamics shown in Figs. 1-2, down to the scale of minutes within a day -- in particular, the huge waves of links which appear like sudden shocks around notable events (Figs. 1c and 1d) and which could further enflame hate, anger and distrust during these events or their aftermath, possibly inciting new violent acts.
(2) They must focus on activity (links) *between* platforms, particularly including the many *smaller* platforms as shown explicitly in Fig. 1d -- and *not* just activity within the largest platforms.

(3) They must avoid chasing down reported links since this resorts to whac-a-mole, i.e. existing links can get buried below later content and hence become less relevant; or get removed on purpose by the community member(s); or the piece of content they are in gets removed. This link loss also means that the most active pathways that hate content spreads though are changing all the time, hence mitigations to prevent system-wide spreading need to account for this.

No current mitigation or legislation accounts for (1)-(3): however, ones that do can easily be developed once we establish governing *dynamical* equations that explicitly include (1)-(3), just as best-choice interventions in an engineering system are based on particular solutions to that system's governing equations[59]. The SI Sec. 5 derives these governing dynamical equations (SI Eq. 102) starting from a realistic online grouping mechanism[60,61]. They reproduce the observation of waves (shockwaves) in link creation (Fig. 1d, Fig. 2) since they are mathematically equivalent to shockwave equations -- even when used in their minimal form (Fig. 2) which is $\dot{S}_i = H(t - t_i)\left[a_i(S_{i,0} - S_i) + \sum_j b_{i,j}(S_j - S_i)\right]$ where $H(.)$ is the Heaviside function, $t_i$ is an expression for the onset time of a new wave of link creation (SI Sec. 5) and most of the $b_{i,j}$ parameters are set to zero for simplicity. These equations are exactly piecewise solvable in their approximate form (SI Sec. 5.5) which means there are no computational errors or instabilities in their solutions and hence predictions.

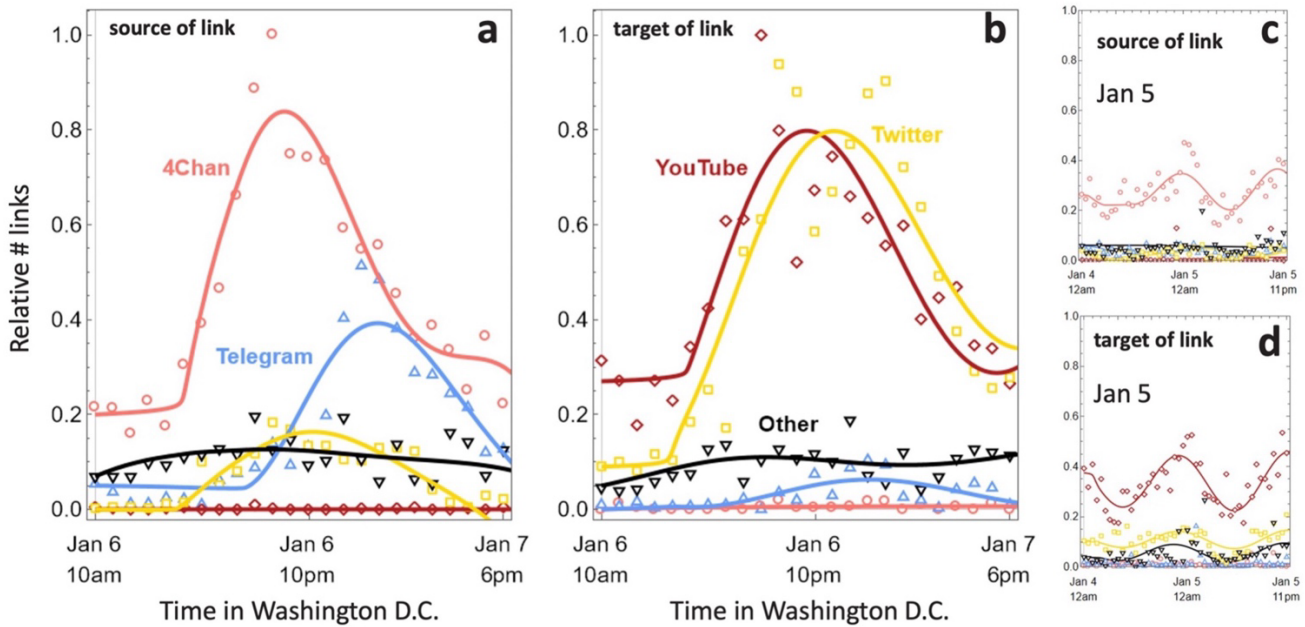

**Figure 2. Empirical data (symbols) vs. mathematical solutions of the deterministic governing equations (curves, derivation SI Sec. 5). a: Relative number of links created at time *t* from hate communities on a given platform (source). b: Relative number of links created at time *t* to other communities on a given platform (target). Approximately 80% of targets are hate-vulnerable mainstream communities. Only the largest curves are shown, the rest are aggregated as 'Other' (black curves). c and d: Same as plots a and b but for Jan 5.**

Figures 2c,d illustrate our finding that the same governing equation solutions that reproduce the link-dynamics for January 6 also reproduce them for more normal days, just with lower amplitudes. This suggests a new interpretation of January 6's evolution: instead of being an unpredictable Black Swan event, it was an amplified version of what was already happening online on January 5. Figure 2 also shows that instead of Twitter being a central driver platform, as assumed in many current discussions, it behaves more like a reporter -- smaller platforms instead act collectively as drivers. Also, the governing equations are deterministic since they operate at the many-link-node level: this means that more general meso- and macro-scale predictions can be made. For example, they can predict the impact of a given mitigation (e.g. on January 5) by forward iteration from that date, and the best mitigation for a given desired impact can be predicted by running this process in reverse.

These findings demonstrate how our analysis provides abiding insights and quantitative predictions -- even down to the hourly scale. Also, these dynamical link equations provide a tipping point condition for whether hate content can spread system-wide or not, and how to prevent it. SI Sec. 6 derives this: here for simplicity we assume a community exposed to new hate content through a link will digest it after time $T_{\text{digest-hate}}$ on average; they will forget it after time $T_{\text{forget-hate}}$ on average; a link that connects clusters of communities appears after time $T_{\text{create-links}}$ on average; and links disappear after time $T_{\text{lose-links}}$ on average. Then Eq. 131 derived in the SI predicts hate content will be prevented from spreading system-wide if $T_{\text{lose-links}} T_{\text{forget-hate}} \left(T_{\text{create-links}} T_{\text{digest-hate}}\right)^{-1} < 1$ which agrees with simulations: hence system-wide spreading will be prevented by increasing the time to create links or digest hate, or decreasing the time to lose links or forget hate, so that this condition is met. This shows mathematically why criteria (1)-(3) are crucial and why mitigation or legislation that does not meet these criteria will not be effective.

Aggregating the link dynamics over time (Fig. 3a) shows even more clearly that online hate does not live at the fringe, and that large platforms are not the key. The many smaller platforms in Fig. 3a act like dynamical glue that binds the hate networks together and attaches them *directly* to the mainstream. Taking the average community size around 10,000 yields a total number of individuals in this battlefield around 6 billion which suggests it captures very crudely the global population. Assuming only 1 in 5 links produce any hate influence, this suggests 1 in 10 of the global population were likely exposed to their influence in recent years. Even if only a tiny fraction then carry out violent acts, say 1 in 100,000, this will generate 10,000 violent acts globally every 2.5 years.

Real-world events involving hate, are mirrored -- and may increasingly be pre-empted -- by hate activity within this dynamical network. In addition to the events in Fig. 1c, Fig. 3a shows the 6 May 2023 Texas shooter's community which had attracted 4 separate links into it from other hate communities prior to his attack. This means that members of these hate communities had been alerted to his YouTube channel and could have easily posted comments and/or content that fueled extreme views and hence influence among his channel's members -- including him. Mentions of RWDS ('Right Wing Death Squad') that appeared as insignia worn by recent mass shooters and members of neo-Nazi units in the Ukraine-Russia conflict, are also prevalent across Fig. 3a -- so too Wagner mercenary communities (see SI Sec. 3). Figure 3a also reveals how some hate-vulnerable communities have far higher exposure risk than others – not necessarily because of their views, but because they are more appealing prey: they attract more links and so sit closer to the hate core in Fig. 3a because of its node-repulsion-link-attraction ForceAtlas2 layout. This also explains its ordered circles akin to solar system orbital structure: successive subsets of hate-vulnerable nodes have 1, 2, 3, etc. links from 4Chan (blue) and hence a net spring force pulling them toward the hate core that is 1, 2, 3, etc. times as strong. These successively smaller radius stripes hence contain hate-vulnerable communities that are roughly 1, 2, 3, etc. times more likely to receive hate content and influence. This suggests tailoring pre-emptive action first on the inner rings closest to the hate core in Fig. 3a, then working in order of increasing radius and hence decreasing risk of exposure.

This new knowledge of the dynamical battlefield and its governing equations, means that the expected impacts of different mitigations can be rigorously calculated and compared for the first time -- and formal control theory then applied[62]. To avoid more math, Fig. 3b illustrates this simply using computer simulations for 6 variants of a mitigation that avoids contentious shutting-down of communities: instead, a post is removed from community A if it contains a link(s) to extreme content (e.g. a hate manifesto or footage from a mass shooting) posted in some other communities B, C etc. Repeating this continually in the limit of slow link dynamics, is equivalent to neutralizing an increasing number of communities from the battlefield in Fig. 3a.

The resulting mitigation impacts over time (Fig. 3b) are unlike prior estimates and again challenge current thinking. Figure 3b inset shows the impact on the relative size $S$ of the largest cluster which, in a slow link-dynamic limit, represents the maximum spread that any piece of hateful content can have. The main panel shows the impacts on the average size $\langle s \rangle$ of the other clusters which quantifies how linked the remaining communities are on average, and hence quantifies the threat they pose as nucleation sites for further activities. The curves are very different to those for exponential or scale-free networks associated with the World Wide Web or Internet as a whole[63]. Also, there is no absolute 'best' mitigation because of side-effects: removing hate nodes irrespective of platform reduces $S$ quicker than removing hate-vulnerable nodes (inset: red curve is lower than blue), but it has the disadvantage that it generates larger nucleation clusters (larger $\langle s \rangle$) for future harms (main: red curve is higher than blue).

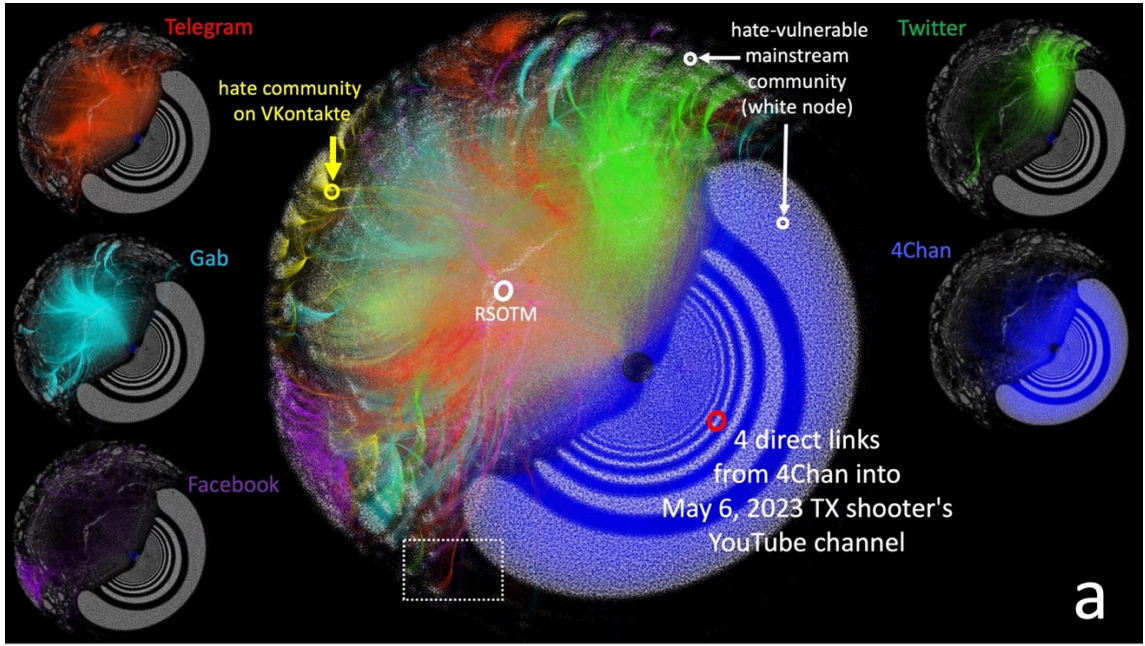

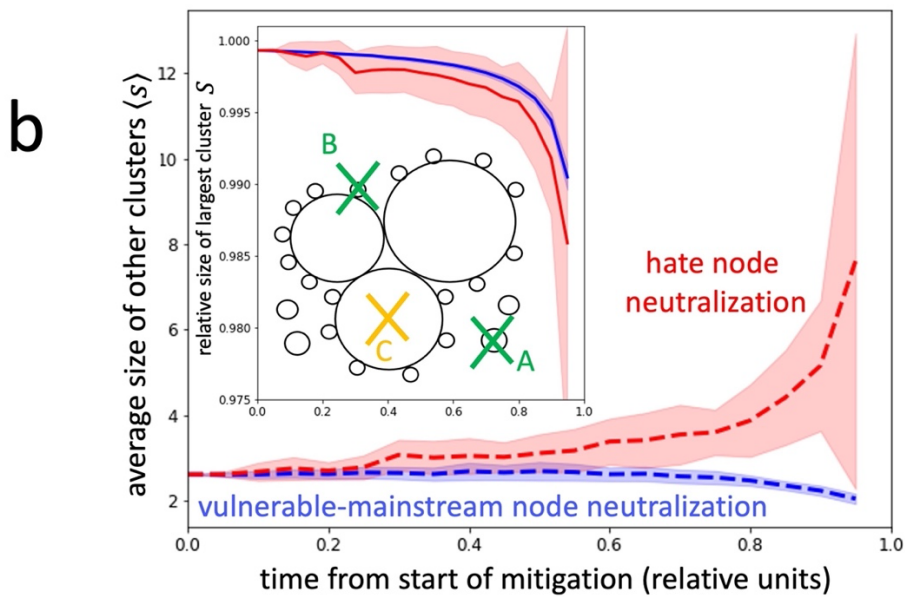

**Figure 3. a: Dynamical battlefield aggregated over previous 2.5 years. Each colored node is a hate community. Each white node is a hate-vulnerable mainstream community to which a hate node has a direct link. Side panels compare platforms' involvement. Alleged 2023 Texas shooter's YouTube community is shown, so too is a major Wagner mercenary community on Telegram (RSOTM, Reverse Side of the Medal). Mentions of RWDS (Right Wing Death Squad) featured in the Texas shooter's chest tattoo, are scattered across the hate core (SI Sec. 3). See SI for others and an example zoom-in (dotted box, SI Sec. 7). Network layout (ForceAtlas2 algorithm[64]) is spontaneous: sets of communities appear closer together when they share more links. b: Comparing impact on battlefield of different mitigation schemes. Time is scaled by the total time to neutralize all communities. Inset uses schematic of the multiple-platform battlefield network to explain these different curve behaviors: neutralizing nodes in clusters like A has little effect on S but decreases ⟨s⟩; neutralizing nodes in clusters like B can slightly decrease S but increase ⟨s⟩; neutralizing nodes in clusters like C can strongly decrease S and strongly increase ⟨s⟩.**

Nobody knows what future AI will be weaponized[65], but knowing the battlefield dynamics clarifies *where* and *how* that war will be fought. Future improvements will include: (1) subclassify hate by type (e.g. anti-Semitic); (2) explore other hate definitions; (3) analyze blends of hate types; (4) sub-classify hate-vulnerable communities; (5) add private platforms; (6) add link weights according their use; (7) add links from mainstream to hate communities; (8) add links to government sources (e.g. Fig. S3); (9) include general harm types; (10) subclassify each community by location or scale (e.g. local vs. global[66,67]). Though our data is technically a large sample of the unknown true online population, the estimated number of many billions suggests it qualifies as a crude population map. Also, we did not obtain the nodes and links by simple sampling but rather by detection and then following links from node to node. This process tended to eventually return to the same nodes and hence, like circling the globe, it hints we have charted out - albeit crudely - the skeleton of the true online hate ecosystem.

**Funding:** N.F.J. is supported by U.S. Air Force Office of Scientific Research awards FA9550-20-1-0382 and FA9550-20-1-0383.

**Authors contributions:** M.Z. generated the figures and mitigation analysis. R.S. collected edge data, managed the databases, and developed software. L.I. analyzed the governing equations and comparison to data. R.S. and N.J.R. were in charge of the data collection and classification. N.F.J. supervised the project and wrote the paper drafts. All authors were involved at some stage in the conceptualization, methodology, analysis and validation, and reviewing the work and drafts.

**Competing interests:** The authors have no competing interests, either financial and/or non-financial, in relation to the work described in this paper.

**All correspondence and material requests should be addressed to** N.F.J. neiljohnson@gwu.edu

**Data and materials availability:** Data that reproduces the figures will be made available online at the authors' website. The code used is Gephi which is free open-source software, and Mathematica which is a well-known commercial product available for free through site licenses in many universities.